\documentclass[article,useAMS,usenatbib]{mn2e}
\usepackage{graphicx}
\usepackage{eufrak}
\usepackage{eucal}
\usepackage{txfonts}
\usepackage{xcolor}

\begin{document}

\title[The sudden appearance of dust in the early Universe]{The sudden appearance of dust in the early Universe}

\author[Mattsson]{Lars Mattsson$^{1}$\\
$^1$Nordita, KTH Royal Institute of Technology \& Stockholm University, Roslagstullsbacken 23, SE-106 91, Stockholm, Sweden\\
}

\date{}

\pagerange{\pageref{firstpage}--\pageref{lastpage}} \pubyear{2015}

\maketitle

\label{firstpage}

\begin{abstract}
Observations suggest that high-redshift galaxies are either very dusty or essentially dust free. The evolution from one regime to the other must also be very fast, since evolved and dusty galaxies show up at redshifts corresponding to a Universe which is only about $500$ Myr old. In the present paper models which predicts the existence of an apparent dichotomy between dusty and dust-free galaxies at high redshift are considered. Galaxies become dusty as soon as they reach an evolved state and the transition is very rapid. A special case suggests that while stellar dust production is overall relatively insignificant -- contrary to what has been argued recently -- it can at the same time be consistent with efficient dust production in supernovae in the local Universe. Special attention will be given to the recent discovery of a dusty normal galaxy (A1689-zD1) at a very high redshift $z=7.5\pm0.2$.  
\end{abstract}


\section{Introduction} 
In the early Universe, when the Universe was less than a few hundred Myr years old (redshifts $z>6$), evidence is lacking of the cold dust seen in the local Universe, except in huge, extreme starburst galaxies (SBGs\footnote{An SBG is here defined as a galaxy that is going or has gone through an extreme phase of star formation, which differs from the empirical definition.}) with large molecular-gas masses of about $10^{10}\,M_{\sun}$ and estimated star-formation rates (SFRs) of a few times $10^3\,M_{\sun}$\,yr$^{-1}$ \citep{Riechers13,Michalowski10,Walter03}. `Normal' galaxies appear to be essentially dust free, with a prototypical  example  being the luminous blue galaxy (LBG) `Himiko'  ($z=6.6$),  for which the estimated star-formation timescale is $\tau_\star  = M_{\rm stars}/{\rm SFR}\sim 150$ \,Myr \citep{Ouchi13,Fisher14}.  Himiko is metal-poor and gas-rich galaxy with a metallicity that is only a few percent (at most) of the solar value $Z_{\sun}$ and essentially no molecules and dust emission detected. An upper limit to the dust-to-stellar mass ratio of $5.0\,10^{-4}$ can be derived, though. It is believed that the  local  starburst galaxy  I~Zw~18,  with  a  metallicity $Z=0.04\,Z_{\sun}$, is a good analogue at low redshift. I~Zw~18 is also essentially dust free, with an estimated dust mass of $450-1800\,M_{\sun}$ and if Himiko has a similar dust-to-gas mass ratio, it would have a dust mass which falls two orders of magnitude below local SBGs of similar stellar mass \citep{Fisher14}. In general, it seems that high-redshift normal galaxies have very little molecular gas and typically no infrared/sub-mm continuum detections \citep{Tan13,Tan14}, which indicate that many of these galaxies are metal-poor and more or less dust-free. Moreover, \citet{Jiang06} have also found that some quasar-host galaxies at high redshift ($z\sim 6$) appear to be essentially without hot dust, which may also be an indication of dust deficiency at low metallicity.

Explaining the high dust masses found in massive, extreme SBGs at high redshifts is a challenge \citep[see, e.g.,][]{Mattsson11}. A simple hypothesis is that stars in these systems form according to a top-heavy initial mass function (IMF) combined with efficient stellar dust production \citep{Gall11,Gall11b}. However, that requires a very low rate of dust destruction due to sputtering and other processes associated with supernova (SN) shockwaves, which is why dust condensation in the interstellar medium (ISM) has been considered to be an crucial ingredient in dust evolution models \citep{Kuo12,Mattsson11}. In fact, interstellar grain growth appears to be needed in the Milky Way too \citep{Zhukovska08} and probably also in other late-type galaxies \citep{Mattsson14b,Mattsson12b,Hirashita99}, which would suggest that condensation in molecular clouds in the ISM is more important than stellar dust production. Yet, detections of large quantities of cold dust in SN remnants \citep[e.g.][]{Matsuura11,Gomez12} are indicative of efficient dust production. The dust-mass estimates are still uncertain and a matter of debate \citep[see e.g.][and references therein]{Mattsson15}, but local SNe are definitely producing significant amounts of dust. If dust production is effectively due to stars it would explain why the dust-to-metals ratio in galaxies seem to show only small variations \citep{Zafar13}, but measurements of dust depletions are suggesting that the overall efficiency of dust production in SNe is either low or increasing with metallicity (De Cia et al. 2013; Mattsson et al. 2014 -- henceforth cited as M14).
 
\citet{Watson15} have recently reported a detection of an evolved, dusty, normal galaxy (A1689-zD1) at redshift $z = 7.5\pm0.2$. This intriguing discovery provides an important piece of the puzzle needed to fully understand the origins of cosmic dust.  A1689-zD1 is not necessarily more challenging to the theory of cosmic dust formation than any previously detected high-redshift galaxies with large dust masses, but it has two distinctive properties which has not been anticipated to appear in combination at that redshift: (1) a very high content of stars, metals and dust; (2) a relatively modest baryon mass (and thus it is not an extreme SBG). 

It has been shown that stellar sources of dust are likely insufficient \citep{Michalowski15}, which opens up for other interpretations. \citet{Mancini15} suggested rapid interstellar condensation due to high-density molecular gas, but that may not apply globally. Thus, in this Letter, the role of the critical {\it metallicity} for efficient condensation is analysed.

%This paper discusses the implications of the fact that galaxies like A1689-zD1 exist when the universe is only $\sim 100$\,Myr old.

\section{Model and hypothesis}
\subsection{Stellar dust and astration/gas-consumption}
The first source of cosmic dust is always the grains expelled by stars. In the high-$z$ Universe, stellar dust production must be dominated by massive relatively short-lived stars. Thus, the recycling time of the cosmic matter cycle can be considered negligible (holds for stars of initial masses $M_{\rm ini} \gtrsim 5 M_{\sun}$) and the timescale for dust injection by stars is $\tau_{\rm stellar}^{-1} = y_{\rm d}\,\rho_{\rm d}^{-1}\, {d\rho_{\star}/ dt}$ where $y_{\rm d}$ is the effective yield, $\rho_\star$ is the stellar-mass density, $\rho_{\rm g}$ is the gas mass density and $\rho_{\rm d}$ is the dust density. But dust is also consumed by stars as they form. The gas-consumption timescale is given by $\tau_{\star}^{-1} = \rho_{\rm g}^{-1}\, {d\rho_{\star}/ dt}$, which means the astration rate is $(d\rho_{\rm d}/dt)_{\rm astr} = \rho _{\rm d}\,\tau_{\star}^{-1}$. How these two timescales compare is crucial for whether stellar dust production can be regarded as efficient or not.

\subsection{The dust condensation and destruction timescales}
The evolution of the condensation timescale is due to two basic facts: the rate of condensation must be proportional to (1) the number density of dust $n_{\rm d} = \rho_{\rm d}\,\langle m_{\rm gr}\rangle^{-1}$, where $\langle m_{\rm gr}\rangle$ is the average grain mass, and (2) the density of available growth species, which introduces a factor $\rho_{\rm g} (Z-Z_{\rm d})$, where $Z_{\rm d} \equiv \rho_{\rm d}/\rho_{\rm g}$, $Z \equiv \rho_{\rm met}/\rho_{\rm g}$. The growth species are of course in the form of molecules, so the molecular fraction should also enter the prescription. To leading order, the {\it local} timescale is then given by
\begin{equation}
  \tau_{\rm cond}^{-1} \equiv {1\over \rho_{\rm d}}\left({d\rho_{\rm d}\over dt} \right)_{\rm cond} \propto  \eta\, \langle m_{\rm gr}\rangle^{-1}(Z-Z_{\rm d})\,{\rho_{\rm g}},
\end{equation}
where $\eta$ is the molecular gas fraction \citep{Mattsson14}. Dimensional analysis shows that in order to obtain a timescale, a factor with unit mass density per time is needed. It is empirically well established that $dM_{\rm stars}/dt \propto M_{\rm mol}$, which justifies to replace $\eta\,\rho_{\rm g}$ with $d\rho_{\star}/dt$ multiplied by some appropriate factor. The {\it global} rate of condensation may thus be expressed as
\begin{equation}
\left({dZ_{\rm d}\over dt} \right)_{\rm cond} \approx Z_{\rm d}\,(Z-Z_{\rm d})\, {\epsilon\over M_{\rm gas}}{dM_{\rm stars}\over dt},\, {\rm with}\,\,\epsilon = \epsilon^{\prime} - {1\over Z-Z_{\rm d}}.
\end{equation}
In case of efficient condensation, i.e., if $\tau_{\rm cond} \ll \tau_\star$, then $\epsilon \sim \epsilon^{\prime} = $\,\,constant, and one arrives at the prescription used below.  

Dust destruction is mainly by sputtering in the high-velocity interstellar shocks driven by SNe. Following \citet{McKee89} the dust destruction time-scale is given by $\tau_{\rm d}^{-1} = \rho_{\rm g}^{-1} \langle m_{\rm ISM}\rangle\,R_{\rm SN}$, where $\langle m_{\rm ISM}\rangle$ is the effective gas mass cleared of dust by each SN event and $R_{\rm SN}$ is the SN rate, which is obviously proportional to $d\rho_\star/dt$ at early times. Thus, the time scale $\tau_{\rm d}$ may be parameterised as $\tau_{\rm d}^{-1} \approx \delta\, \rho_{\rm g}^{-1} \,{d\rho_{\rm s}/ dt}$, where $\delta$ is the dimensionless dust destruction parameter. However, small grains are more susceptible to destruction by sputtering in SN shock waves than large grains \citep{Slavin04}. As discussed in \citet{Mattsson14}, the timescale of dust destruction may therefore not only be inversely proportional to the SN rate, but also to the abundance of dust. An adequate modification to the dust-destruction timescale would then be to introduce a dependence on the dust-to-gas ratio $Z_{\rm d}$, i.e.,
\begin{equation}
\label{taud2}
\tau_{\rm d}^{-1} \approx  {Z_{\rm d}\over Z_{\rm d,\,G}}{\delta \over \rho_{\rm g}}{d\rho_{\rm s}\over dt},
\end{equation}
where $Z_{\rm d,\,G}$ is the present-day Galactic dust-to-gas ratio. 

\subsection{Timescale comparisons}
%Defining the dust-to-stellar mass ratio as $M_{\rm d}/M_{\rm stars}\equiv \rho_{\rm d}/\rho_\star$ we can express the log-log slope of the dust-mass density $\rho_{\rm d}$ as a function of the stellar-mass density $\rho_\star$ as
%\begin{equation}
%\alpha \equiv {d\ln \rho_{\rm d}\over d\ln \rho_\star} = {1\over M_{\rm d}/M_{\rm stars}}{d\rho_{\rm d}\over d\rho_\star}.
%\end{equation}
The derivative $d\rho_{\rm d}/ d\rho_\star$ represents the dust-production rate relative to the growth rate of the stellar component. If $d\rho_{\rm d}/ d\rho_\star$ is split into sources and sinks, one arrives at the equation
\begin{equation}
\label{alpha}
{d\rho_{\rm d}\over d\rho_\star} = \left({d\rho_{\rm d}\over d\rho_\star}\right)_{\rm stellar} + \left({d\rho_{\rm d}\over d\rho_\star}\right)_{\rm cond} - \left({d\rho_{\rm d}\over d\rho_\star}\right)_{\rm destr} - \left({d\rho_{\rm d}\over d\rho_\star}\right)_{\rm astr},
\end{equation}
where each of the terms on the right-hand side correspond to stellar (supernova) production, condensation in the ISM, destruction (by sputtering) in the ISM and astration of dust, respectively.

If the net stellar dust production is negligible, i.e., if the first and the last term more or less cancel, the formation of the interstellar dust component will be determined by what happens in the ISM and how the condensation timescale compares to the destruction timescale. (Such a scenario will still correspond to substantial stellar dust production at later stages, in accordance with the dusty SN remnants found locally.) As a consequence, the dust fraction would be expected to make a more or less sudden jump from a small initial dust fraction (presumably suppled by an early generation of stars) to a high degree of dust condensation once a critical metallicity in the ISM is reached, provided that condensation in the ISM is efficient. The existence of such a critical metallicity\footnote{More precisely, there is critical density of the relevant growth species, which is usually obtained within a narrow range of metallicities.} is  well established and should occur just above $Z\sim 0.1\,Z_{\sun}$ for a normal galaxy \citep{Asano13,Kuo12,Mattsson12} and may be the fundamental reason why there are both dust-free and dust-rich young SBGs in the local Universe and likely also at high redshifts. It is therefore of utmost importance to investigate a scenario where stellar dust production and gas consumption have similar timescales throughout the course of evolution of the ISM. In such case, the evolution of the dust mass is effectively a result of the processing in the ISM, i.e., 
\begin{equation}
\left({d\rho_{\rm d}\over d\rho_\star}\right)_{\rm stellar} \approx \left({d\rho_{\rm d}\over d\rho_\star}\right)_{\rm astr}   \rightarrow
{d\rho_{\rm d}\over d\rho_\star} \approx \left({d\rho_{\rm d}\over d\rho_\star}\right)_{\rm cond} - \left({d\rho_{\rm d}\over d\rho_\star}\right)_{\rm destr}.
\end{equation}
Assuming instantaneous recycling, this situation would occur if $y_{\rm d} \approx Z_{\rm d}$, which is not to say that $y_{\rm d}$ actually depend on $Z_{\rm d}$ (but $y_{\rm d}$ is indeed a function of time).

\subsection{Modelling early dust formation}
Assuming rapid galaxy formation, the `closed box' model is a valid approximation for the enrichment of metals.
Adopting a `closed box' and the instantaneous recycling approximation, the evolution of the dust-to-gas ratio $Z_{\rm d}$ is governed by
\begin{equation}
\label{dtm}
{dZ_{\rm d}\over dZ} =  {y_{\rm d}\over y_Z} + {Z_{\rm d}\over y_Z}\left[\epsilon \left(1-{Z_{\rm d}\over Z} \right)\,Z -\delta^{\prime} Z_{\rm d}\right],
\end{equation}
where $\delta^{\prime} = \delta\,Z_{\rm d,\,G}^{-1}$ and $y_{\rm d}$, $y_Z$ are the effective stellar dust and metal yields, respectively. The timescales for dust condensation and dust destruction in the ISM are defined as in the `equilibrium model' in \citet{Mattsson14} and parameterised above by $\epsilon$ and $\delta^{\prime}$, respectively. 

In very young objects all stellar dust production (dust injection to the ISM) must be due to SNe/massive stars, since low and intermediate stars cannot have had time to evolve. The importance of massive stars as dust producers at early times is unclear. In the local Universe supernovae seem to produce large amounts of dust \citep[see, e.g.][]{Matsuura11,Gomez12}, but estimates of dust depletion at different metallicities seem to indicate that early stellar dust production must have been much less efficient \citep{DeCia13,Mattsson14}. It is therefore reasonable to consider three scenarios: (1) a completely supernova-dominated scenario where $y_{\rm d}\sim y_Z$; (2) a condensation-destruction equilibrium model of the type discussed in \citet{Mattsson14} with very limited supernova-dust production ($y_{\rm d}= 10^{-5}$), but efficient dust condensation in the ISM; (3) a model similar to scenario 2, but with an ISM-dominated evolution where it is assumed that $\tau_{\rm stellar} = \tau_{\star}$, which corresponds to $y_{\rm d} = Z_{\rm d}$. A very substantial part of all dust production is still due to condensation in the ISM, but stars will contribute significantly at later stages. Observational evidence suggest the rate of dust destruction in the ISM must be low, since a very high degree of dust condensation would not otherwise be possible. The empirical conclusion is therefore that at early times, dust destruction is inefficient, which is consistent with the prescription for the condensation timescale used to arrive at eq. \ref{dtm} above \citep[see][for further details]{Mattsson14}. The `equilibrium assumption' $\delta^{\prime} = \epsilon$, leading to a dust-to-metals ratio of $1/2$, will be used in the following.

\begin{itemize}
\item {\bf Scenario (1)} where the dust to metals ratio is a constant assumed to be $y_{\rm d}/y_Z = 0.5$, corresponds to having 50\% of all metals in dust to match the local dust-to-metals ratio, which seems to be almost universal for evolved galaxies. The model corresponds to a high dust-to-metals ratio at all times and can therefore not explain possibly dust-free galaxies at high redshift, however.

\item {\bf Scenario (2)} yields a solution identical to eq. (20) in \citet{Mattsson14} with $\epsilon = \delta^{\prime}$. The reader is referred to that paper for further details. With the stellar dust yield taken to be just $y_{\rm d}= 10^{-5}$ (to match the low $M_{\rm d}/M_{\rm stars}$ of I~Zw~18), almost all of the dust mass has to be due to dust condensation in the ISM. This model does indeed rapidly evolve from an almost dust-free to a dusty regime, but cannot account for the high dust-formation efficient in SNe suggested by observations. This model is included as an example of a model which agrees quantitatively with the dust masses in high-$z$ SBGs, despite being qualitatively incorrect.

\item {\bf Scenario (3)} is similar to scenario 2, but letting $y_{\rm d} = Z_{\rm d}$ yields a solution of the form,
\begin{equation}
Z_{\rm d}(\tilde{Z}) = \left\{ 2{\epsilon + \delta^{\prime}\over \gamma}\left[F\left({1\over \gamma}{d\xi\over d\tilde{Z}} \right) - e^{\,\xi_0-\xi}F\left({1\over \gamma}{d\xi_0\over d\tilde{Z_0}} \right) \right] + {e^{\,\xi_0-\xi}\over Z_{\rm d}(0)} \right\}^{-1},
\end{equation}
where $F(z)$ is a Dawson function with argument $z$, $\gamma \equiv \sqrt{2\,y_Z\,\epsilon}$, $\xi = \tilde{Z}+\gamma^2/4\,\tilde{Z}^2$, $\xi_0 = \tilde{Z}_0+\gamma^2/4\,\tilde{Z}_0^2$ and $\tilde{Z} \equiv Z/y_Z$, $\tilde{Z}_0 \equiv Z_0/y_Z$. To obtain the dust/no-dust dichotomy, the initial dust-to-gas mass ratio $Z_{\rm d}$ should be roughly 1\% of the initial metallicity $Z_0$, which must be orders of magnitude smaller than the final metallicity. 
%The assumed equality of the stellar dust yield to the dust-to-gas mass ratio ($y_{\rm d} = Z_{\rm d}$) can be interpreted in two ways: either as spurious correlation that one may use to construct the model, or, as a simple description of a somewhat speculative scenario where condensation in supernova remnants requires pre-existence of some condensed material to act as seed grains. In the latter case, interstellar dust existing around the star prior to the supernova explosion, must survive to some extent\footnote{Here one may note that also massive stars can have a lot of rocky material encircling them, which need not be totally disintegrated in a SN explosion.}. 
\end{itemize}

      \begin{figure*}
   % \sidecaption
  \resizebox{\hsize}{!}{ 
  \includegraphics{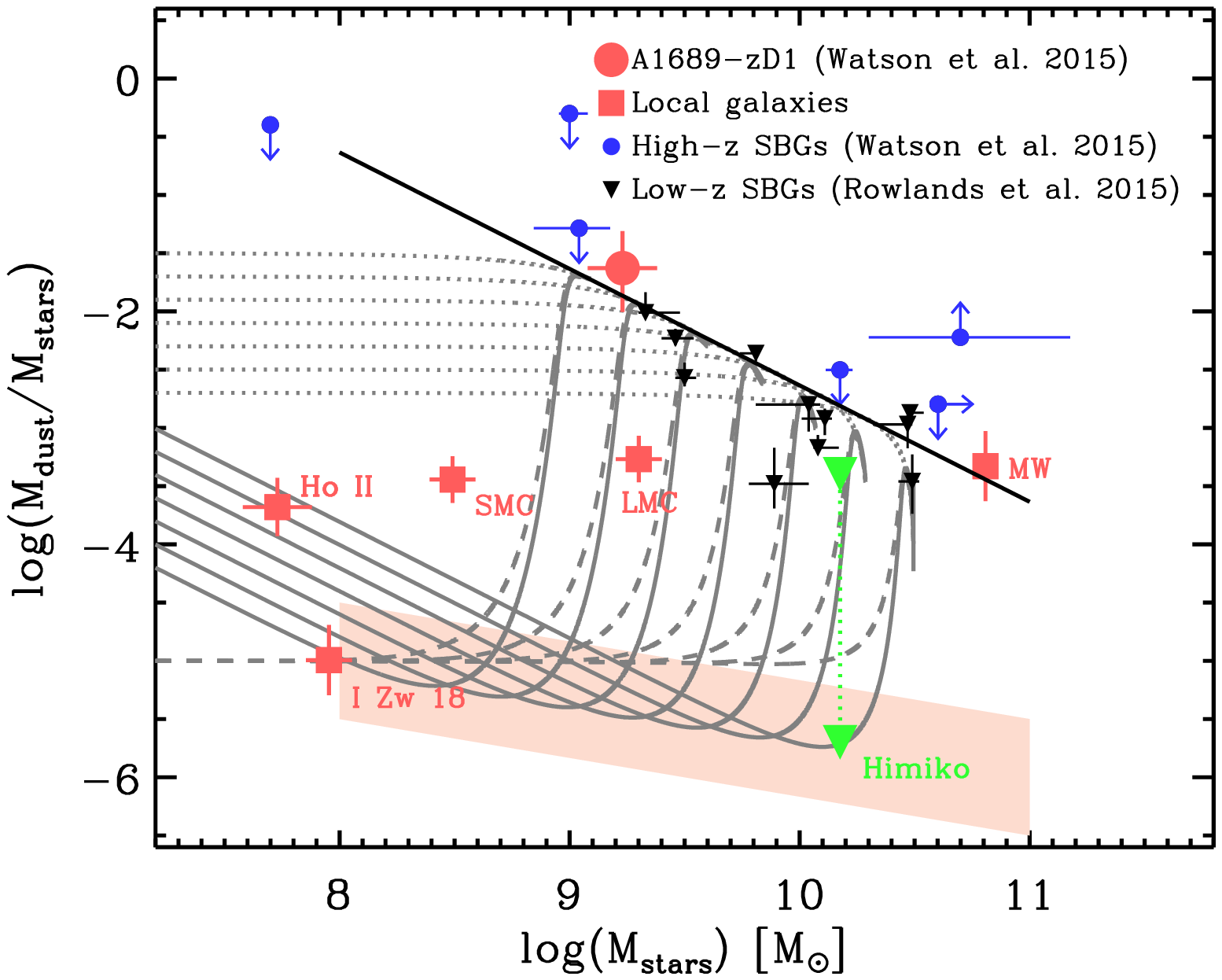}
  \includegraphics{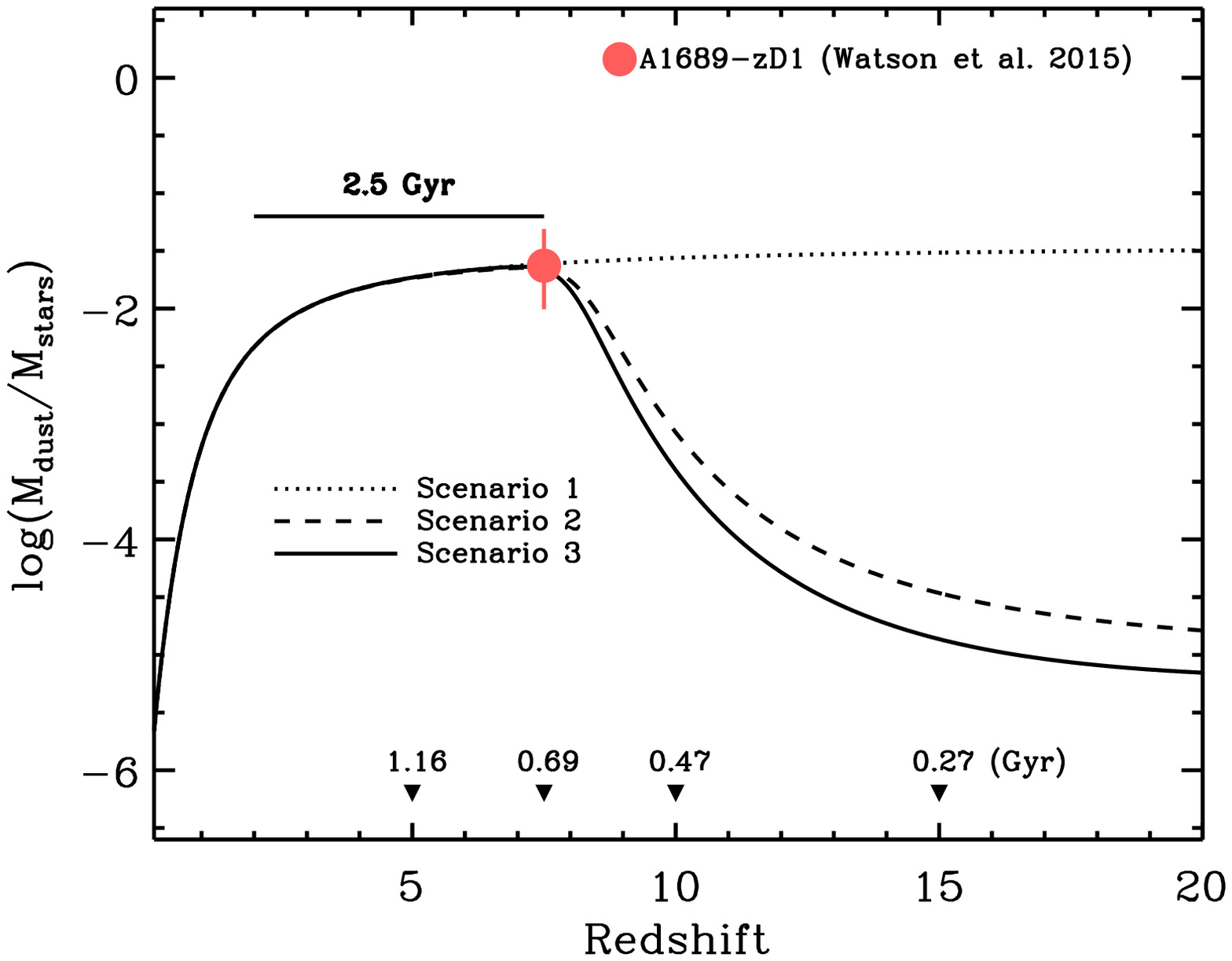}}
  \caption{Left panel: models (grey lines) for different baryon masses compared to observed dust and stellar masses for local as well as high redshift objects. Dotted lines correspond to scenario 1, dashed lines to scenario 2 and solid lines to scenario 3 (see text for details). The solid black line shows the $\alpha = 0$ case for $M_{\rm d}\approx 10^7\,M_{\sun}$. For the $z=6.6$ galaxy `Himiko', the plotted range in dust mass is based on the observational upper limit (no detection) and the expected dust mass using I Zw 18 as template. The shaded region shows where the metal-poor, dust-free galaxies should appear in the diagram if such small amounts of dust could be detected. Right panel: best fit models for A1689-zD1 ($z=7.5$) assuming a baryon mass $M_{\rm tot} = 3.7\,10^9\,M_{\sun}$ and gas-consumption timescale $\tau_\star = 1.0$\,Gyr. The numbers above the x-axis indicate the age of the Universe in Gyr at redshifts $z = 5,7.5,10,15$, assuming a canonical $\Lambda$CDM Universe.}
  \label{dts}
  \end{figure*}

\section{Results and discussion}
\subsection{A universal dust mass and a top-heavy IMF?}
Given the results in previous section, the first thing to consider might be where A1689-zD1 end up in the $M_{\rm d}/M_{\rm stars}$ vs. $M_{\rm stars}$ diagram. As shown in Fig. \ref{dts} (left panel), A1689-zD1 follows the trend established by other star-forming galaxies/SBGs at both high and low redshift \citep[data taken from][]{Watson15, Rowlands15} -- a trend which corresponds to a common dust mass of $M_{\rm d}\sim 10^7\,M_{\sun}$. Thus, the dust mass of A1689-zD1 and other galaxies at high redshift is not peculiar in relation to the stellar mass, simply because it is largely uncorrelated with the stellar mass. Note, however, that local star-forming dwarf galaxies fall below this trend, according to recent estimates of their dust masses \citep{Skibba11,Skibba12}. The $M_{\rm d}/M_{\rm stars}$ value of the Milky Way \citep[data taken from table 1 in][]{Riechers13} is similar to the local dwarf galaxies, but falls on the $M_{\rm d}/M_{\rm stars} \propto 1/M_{\rm stars}$ trend because of the stellar mass.

The overall trend corresponds to $dM_{\rm d}/dM_{\rm stars} = 0$ (shown by the black line in the left panel of Fig. \ref{dts}), which may appear as evidence of a passive dust evolution along that trend once the galaxies have reach an evolved state and become heavily enriched with dust. This is not likely the case, however. First, the trend exists over almost two orders of magnitude in stellar mass and evolved galaxies cannot move much in the $M_{\rm d}/M_{\rm stars}$ vs. $M_{\rm stars}$ diagram since they would have converted at least half of their gas into stars. (A doubling of the stellar mass corresponds to $\sim 0.3$\,dex shift.) Second, keeping $M_{\rm d}$ close to constant over an extended period of time (at least 1\,Gyr), should coincide with a low SFR. Hence, the trend seen in Fig. \ref{dts} is rather an effect of galaxies of different masses having different effective metal yields, i.e., $y_Z$ is decreasing with galaxy mass, and the fact that the galaxies are observed during a phase when their dust masses are close to their maxima. This phase is also relatively long compared to the almost dust-free phase. In the models constructed specifically for A1689-zD1 (see Fig. \ref{dts}, right panel), the time between $z = 7.5$ and the point when the dust-to-stellar mass ratio $M_{\rm d}/M_{\rm stars}$ starts to decline rapidly is about 2.5\,Gyr. This should be compared to the 0.7\,Gyr that had passed from the onset of reionisation ($z\approx 20$) until the time of observation ($z = 7.5$)\footnote{Assuming a $\Lambda$CDM universe with DM, dark energy and baryonic density-parameter values of ($\Omega_{\rm m}, \Omega_\lambda, \Omega_{\rm b})=(0.3, 0.7, 0.04)$ and a Hubble constant $H_0=70$~km~s$^{-1}$~Mpc$^{-1}$.}.

All models shown in Fig. \ref{dts} are based on the assumption $y_Z = 1.26\times 10^8\,\left({M_{\rm tot}/ 1\,M_{\sun}}\right)^{-1}$. 
This ansatz is motivated by the fact that evolved galaxies seem to have an almost universal dust-to-metals ratio of $\sim 0.5$, although this constancy will break down for unevolved galaxies at low metalicity \citep{Mattsson14,DeCia13,Zafar13}. Thus, the dust mass for evolved galaxies is $M_{\rm d} \approx 0.5\,Z\,M_{\rm gas}$, which implies that the mass of metals $Z\,M_{\rm gas}$ is roughly a universal constant too for evolved systems, because in such case $M_{\rm d}\sim $ constant. By taylor expansion one can easily show that for a slowly varying metallicity $Z\ll 1$, a yield $y_Z \propto 1/M_{\rm tot}$ corresponds to $Z\,M_{\rm gas} = M_{\rm d} =\,$constant to first order. 

\subsection{Model results}
In scenario 1, that was discussed above, the dust simply follows the metal production. Hence, if the metals are there, so is the dust and there would be no problem having young dusty and metal-rich galaxies at any redshift as long as they has formed enough stars to supply the metals. However, this simplistic model will fail if not all young galaxies at high redshift have relatively high dust-to-stellar mass ratios (see grey dotted lines in the left panel of Fig. \ref{dts}). If I Zw 18 is prototypical for very metal-poor SBGs in general and such systems can be found in the early Universe too (e.g., the galaxy Himiko at redshift $z = 6.6$), then SBGs cannot have a constant dust-to-metals ratio at all metallicities either.

The problem with scenario 1 can in principle be solved by adopting scenario 2. If the effective stellar dust yield is very small (recall that here $y_{\rm d} = 10^{-5}$ is assumed), low dust-to-stellar mass ratios like that in I Zw 18 will be inevitable in metal-poor galaxies (see grey dashed lines in Fig. \ref{dts}). If  $y_Z \propto 1/M_{\rm tot}$, this model can also reproduce the $M_{\rm stars}^{-1}$ trend, since dust condensation in the ISM will set in once the critical metallicity is reached and compensate for the lack of stellar dust production. Following \citet{Mattsson14}, the models shown assume $\epsilon = \delta^{\prime} = 7.5\cdot 10^2$, which corresponds to rapid dust condensation and matches the expected dust-destruction timescale in the Milky Way today \citep[see section 4.2 in][]{Mattsson14}, i.e., efficient destruction in the ISM at later times . 
However, there is a major problem with scenario 2 as well. Recent detections of cold dust in SNRs \citep[e.g.,][]{Matsuura11,Gomez12} show that, at least in the local Universe, supernovae must be efficient dust producers. \citet{Mattsson14} suggested an effective stellar dust yield which depends on metallicity such that there is a transition from inefficient to efficient stellar dust production at some point. But this effect cannot easily be disentangled from the effects of the evolutionary histories of the galaxies. Many local galaxies have had a formation scenario which is incompatible with a `closed box', so the effective metallicity dependence may actually include effects of baryonic infall.

In Scenario 3, where the effective stellar dust yield is $y_{\rm d} = Z_{\rm d}$, a more or less sudden occurrence of dust can be achieved without being at odds with the SN dust production observed in the present-day Universe. Since the astration of dust is exactly cancelled by the stellar dust production in this model, the production and destruction of dust is effectively due to processes in the ISM only. Because $y_{\rm d} = Z_{\rm d}$, the natural initial conditions $Z_{\rm d}(0) = 0$, $Z(0) = 0$ will be problematic. However, assuming that the initial conditions are set by some very brief initial phase of stellar dust and metals production (e.g. from Population III stars) the initial conditions may satisfy  $1\gg Z_{\rm d}(0) > 0$, $1\gg Z(0) > 0$. In the models shown in Fig. \ref{dts} (solid grey lines), it has been assumed that $Z_{\rm d}(0) = 5.0\cdot10^{-7}$ and $Z(0) = 5.0\cdot10^{-5}$, which corresponds to a dust-to-metals ratio of just 1\%, consistent with what is found for I Zw 18 \citep{Fisher14,Herrera-Camus12}. The dust production is very rapid during the most intense phase of dust condensation. The transition from essentially dust free to highly dust enriched is so fast (the expected timescale for A1687-zD1 is $\sim 0.1$\,Gyr) that transitional objects should probably be difficult to find observationally. Effectively, the model then predicts an apparent  dichotomy: a galaxy will either be dust-free or very dusty. This is not fundamentally different from scenario 2, but the jump is slightly faster and, as mentioned above, the model does not violate the observational fact that local SNe produce dust. The shaded area in the left panel of Fig. \ref{dts} shows where early metal-poor galaxies are expected to be found \citep{Tan13}, which is also where the LBG Himiko is likely located, although the non-detection of dust results in an upper limit that is similar to the $M_{\rm d}/M_{\rm stars}$ values of the other SBGs. However, in scenario 3, galaxies seem to naturally evolve towards this anticipated sequence in Fig. \ref{dts}.

The effective stellar dust yield required to explain A1689-zD1 with only stellar dust is $y_{\rm d} \ge 0.035$. This value suggests a very high degree of dust condensation since the required metal yield is similar, assuming that A1689-zD1 has roughly solar metallicity. Moreover, it suggests the efficiency of metal production should be higher than in, e.g., the Milky Way and because of the $M_{\rm stars}^{-1}$ trend seen in Fig. \ref{dts} the efficiency should decrease with galaxy mass, as discusses above. There can be two main reasons for this: (1) the effective IMF is more top-heavy in less massive star-forming galaxies; (2) the accretion (infall) of gas continues for a longer time in more massive galaxies, which will appear as a smaller effective yield if assumed to be closed boxes \citep{Edmunds90}. The latter reason is probably the less important one, since it can only amount to approximately a factor of two. Also, the fact that the amount of metals relative to the stellar mass is smaller in, e.g., the LMC, which has a total mass similar to A1689-zD1, is indicative of a difference in the absolute efficiency of metal production. Such a situation could occur if the IMF is biased towards formation of massive stars. A `top-heavy IMF' to explain early SBGs has indeed been suggested in previous work \citep[e.g.,][]{Gall11,Gall11b} and may also be of importance for the early chemical evolution of the Milky Way \citep[see, e.g.,][]{Mattsson10}, which is why this explanation is not too far-fetched. Note, however, that the effective yield is likely not varying much for local galaxies, because $M_{\rm d}/M_{\rm stars}$ is not showing much variation with stellar mass.

Despite the need for a high metal yield, A1689-zD1 does not seem to have a peculiar evolution, though. As shown in the right panel of Fig. \ref{dts}, models assuming an e-folding gas-consumption timescale of $\tau_\star \sim 1$\,Gyr  will reproduce the dust-to-stellar mass ratio at the correct redshift. A formation timescale of about 1\,Gyr is what is expected for the early formation phase of the Milky Way \citep[see, e.g.,][and references therein]{Mattsson10}, when the halo and the thick disc is expected to have formed, which suggests the evolutionary rate of A1689-zD1 is not extreme.

\section{Conclusions}
Two of the considered scenarios predict the coexistence of two distinctly different types of galaxies at high redshift: the very dusty galaxies and the essentially dust-free galaxies. In these models galaxies will be dusty as soon as they begin to reach an evolved state (with the transition phase from `unevolved' to `evolved' defined as the phase when $M_{\rm gas}/M_{\rm tot} \sim 0.5$) and the transition takes place on a Myr timescale, i.e., it is very rapid. This is in accordance with recent results indicating rapid evolution at high redshift and that large dust masses appear as soon as there is enough metals to form dust efficiently. A far-reaching consequence is that stellar dust production is relatively insignificant, contrary to what has been suggested based on recent discoveries of large cold-dust masses in supernova remnants \citep[e.g.,][]{Matsuura11}. However, scenario 3 is at the same time consistent with highly efficient dust production in supernovae in the local Universe since the stellar dust yield increases with the dust-to-gas mass ratio. 

In summary, it has been suggested that:
\begin{enumerate}
\item Galaxies at high redshift are either dusty or dust-free, where the latter type is statistically underrepresented due to the relatively short time (a few Myr at most) spent in that phase.
\item Galaxies at high redshift stay dusty for a considerable time after their rapid transition from essentially dust-free to dusty.
\item Less massive galaxies must produce more metals per generation of stars, indicating a top-heavy IMF, compared to more massive galaxies. 
\item There is a `universal dust mass', i.e., both local and high-redshift SBGs have very similar dust masses irrespective of their total/stellar masses, which is probably a consequence of (ii) and (iii) in combination.
\item A1689-zD1 is a galaxy with normal evolution, but likely with a very top-heavy IMF.
\end{enumerate}

As a final remark, one may note the following: the dust content of galaxies depends on essentially just one parameter -- the metallicity -- which in turn is mostly a reflection of how much star formation a galaxy has undergone in the past, regardless of its redshift. These galaxies may not be so special after all, but only at high redshifts may two seemingly similar galaxies have vastly different dust content, because the rapid formation and evolution enables a very sudden appearance of dust once a critical metallicity is reached.
 
\section*{Acknowledgments}
A. Andersen and D. Watson are thanked for helpful comments and discussions.
Nordita is funded by the Nordic Council of Ministers, the Swedish Research Council, and the two host universities, the Royal Institute of Technology (KTH) and Stockholm University.

\end{document}